\documentclass[vecphys]{svmult}

\usepackage{makeidx}         
\usepackage{graphicx}        
\usepackage{multicol}        
\usepackage[bottom]{footmisc}
\usepackage[latin1]{inputenc}

\makeindex             

\begin{document}

\title*{The origin of $\sigma$--drops: mapping stellar kinematics and populations in spirals}
\titlerunning{The origin of $\sigma$--drops}
\author{Eric Emsellem\inst{1}}
\institute{Université de Lyon 1, CRAL, Observatoire
de Lyon, 9 av. Charles Andr\'e, F-69230 Saint-Genis Laval; CNRS, UMR 5574 ; ENS de Lyon, France 
\texttt{emsellem@obs.univ-lyon1.fr}}
%
%
\maketitle

\section{Introduction}
\label{sec:1}
The link between nuclear activity and the host galaxy remains elusive. It seems
now clear that the galactic environment is not the key, and that there is no
significant apparent difference in terms of the presence of bars/spirals between
active and non-active galaxies. Nuclear activity nevertheless requires a small
central gas reservoir ($10^4$-$10^7$~M$_{\odot}$) which may be only partly
consumed during the $\sim 10^6-10^7$~yr AGN duty cycle. A possible route
towards a better understanding of the involved physical processes is to directly
probe the gravitational potential of the host galaxies, with the hope that the
corresponding sensitive tracers (stars, gas) will deliver convincing evidence
for a kinematic signature associated with the AGN. This is the motivation for an
on-going survey of 50 galaxies, for which the two-dimensional kinematics of the
(neutral, ionised and molecular) gas and stellar components are being obtained
(see Dumas et al., this conference). With only 10 to 15\% of Seyfert galaxies in
the local Universe, we must also acknowledge the possibility that this activity
is a recurrent (but short) process, and that signatures of recent gas accretion
associated with the onset of the central activity should be detectable. This
paper is a short report on the detection and study of $\sigma$--drops which we
believe are the result of past gas accretion followed by subsequent star
formation.

\section{$\sigma$-drops: the list is growing}
\label{sec:3}
In the course of an observational program aimed at revealing dynamical evidences
for the presence of inner bars in spiral galaxies, we have detected a decrease
of the stellar velocity dispersion towards the centre of 3 out of 3 barred
galaxies \cite{Debca1} where we could probe the very central region. We
suggested that these $\sigma$--drops are the result of gas accretion followed by
star formation, and therefore indicative of a relatively recent infall of
dissipative material. This idea was successfully tested for the specific case of
an infall driven by a large-scale bar \cite{Debca2}, although this does not mean
that bars are the preferred cause for such a process.

\begin{figure}
\centering
\includegraphics[width=12.3cm]{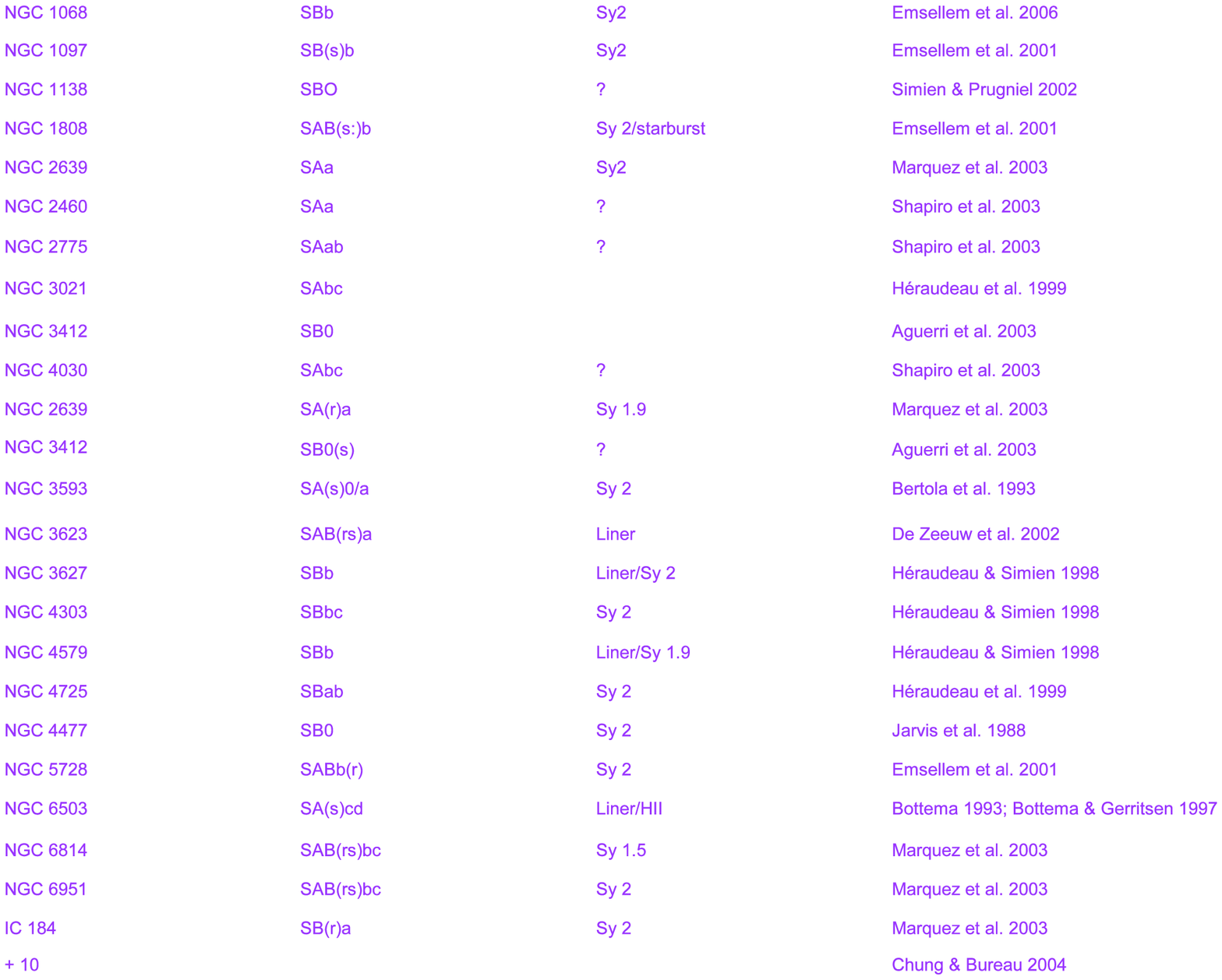}
\caption{A (non-exhaustive) list of galaxies exhibiting a $\sigma$--drop with associated references. The columns are, from left to right: name of the galaxy, Hubble type, activity type and reference.}
\label{fig:list}      
\end{figure}

Since this study of a very small sample of barred galaxies, the list of $\sigma$--drops has
been growing very significantly (and steadily). Fig~\ref{fig:list} provides a non-exhaustive
list of such objects and shows that $\sigma$--drops are indeed a rather common
phenomenon. Most of the galaxies listed in Fig~\ref{fig:list} contain active
nuclei, but this is most certainly due to a bias in the observed sample
where reliable stellar kinematics have been obtained.

\section{Formation scenarios}
\label{sec:4}
\begin{figure}
\centering
\includegraphics[width=12.3cm]{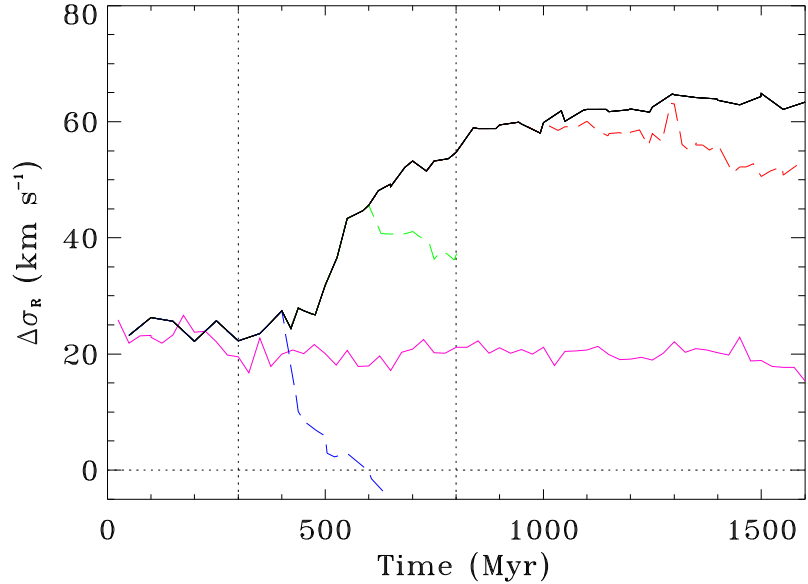}
\caption{Amplitude of the $\sigma$--drop in numerical simulations of spiral
   galaxies, where central star formation is sustained via regular gas infall
      (black curve), or stopped at different times (red, green, blue). The
      magenta curve corresponds to a similar simulation but where star formation
      is not turned on. Extracted from \cite{WC06}.}
\label{fig:wc06}      
\end{figure}

Numerical simulations including stars, gas and star formation are very helpful
in constraining the formation process of $\sigma$--drops in the central regions
of disc galaxies. Recent experiments \cite{WC06} clearly showed that
$\sigma$--drops can be formed on a relatively short timescale (500~Myr) and are
long-lived if star formation is sustained at a level of about 1~M$_{\odot}$/yr
in the nuclear region. This easily occurs in a gas-rich spiral galaxy where a
bar can funneled gas towards the central 500~pc or so. When star formation
stops, the amplitude of the $\sigma$--drop starts to significantly decrease, as
shown in Fig.~\ref{fig:wc06} \cite{WC06}.

\section{A 3D view}
\label{sec:5}
\begin{figure}
\centering
\includegraphics[width=12.3cm]{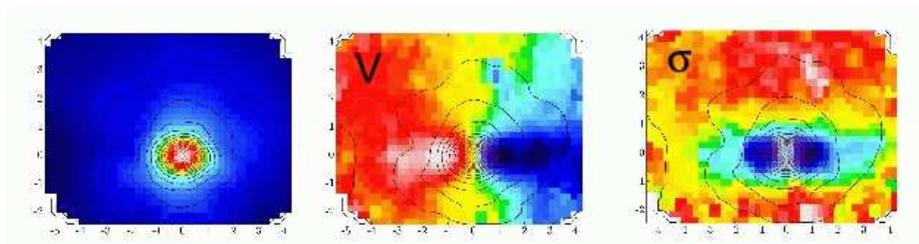}
\caption{VLT/FLAMES maps of the galaxy NGC\,3623. From left to right:
   reconstructed image, stellar velocity field and stellar velocity dispersion
      field. The $\sigma$-drop is obvious and associated with the pinching of
      the (disk-like) isovelocities. }
\label{fig:flames}      
\end{figure}

Integral-field spectrographs (IFS) are now routinely used in order to probe the
full extent and shape of $\sigma$--drops, and hence to better understand the
stellar component with which they are associated. Recent IFS data obtained on
the prototypical Seyfert~2 galaxy NGC\,1068 clearly demonstrated how critical is
such a detailed mapping of the kinematics in the central region
\cite{N1068,Gerssen+06}. Surveys of spiral galaxies using integral-field
spectrographs \cite{Ganda+06} are now showing that
about one third of spirals are expected to exhibit a $\sigma$--drop, this being
a lower limit considering the difficulty of such a detection when dust is at play.
This significant fraction of galaxies with $\sigma$--drops would suggest the
presence of relatively young stars associated with the underlying dynamically
cold stellar component. However, there seems to be no direct relation between the
presence of a $\sigma$-drop and the mean luminosity weighted age of the stellar
population. In order to further test this issue, we have obtained high spectral resolution ($R=12000$)
integral-field data of a known $\sigma$--drop galaxy, NGC\,3623, using FLAMES at the VLT.
Fig.~\ref{fig:flames} shows the corresponding two-dimensional stellar kinematics for this galaxy, 
emphasizing the presence of the central $\sigma$--drop in this galaxy. A preliminary
analysis of the stellar populations in that region shows that stars associated
with the $\sigma$--drop cannot be younger than 1~Gyr, confirming the fact that $\sigma$--drops
can be long-lived.

\section{Conclusions}
\label{sec:6}
$\sigma$-drops are ubiquitous in spirals, with a relative fraction of about 30\%
or more. This suggests that significant episodes of star formation occur
recurrently in the central part of these galaxies. This could in fact be linked
with the common presence of blue nuclei in late-type galaxies
\cite{Rossa+06}. The detection of a young stellar component associated
with a $\sigma$-drop is rather difficult and requires high spectral resolution
and high signal-to-noise ratio data. We have recently started a more systematic
study of these stellar populations using the near-infrared integral-field
spectrograph SINFONI. The analysis of these data is on-going, and should help us
to constrain the star formation history in the central few hundred parcsecs.
This may eventually lead to a better understanding of the
chronology/synchronicity between star formation and activity in the nuclear
region.
%
%



\printindex
\end{document}